\documentstyle[prl,twocolumn,epsf,aps]{revtex}
\begin{document}
\draft

\twocolumn[\hsize\textwidth\columnwidth\hsize\csname @twocolumnfalse\endcsname

\title{Critical temperature and the transition from quantum to classical 
order parameter fluctuations in the three-dimensional Heisenberg 
antiferromagnet}

\author{Anders W. Sandvik}
\address{Department of Physics, University of Illinois at Urbana-Champaign,
1110 West Green Street, Urbana, Illinois 61801}

\date{\today}

\maketitle

\begin{abstract}
We present results of extensive
quantum Monte Carlo simulations of the three-dimensional
(3D) $S=1/2$ Heisenberg antiferromagnet. Finite-size scaling of the spin 
stiffness and the sublattice magnetization gives the critical temperature 
$T_{\rm c}/J = 0.946 \pm 0.001$. The critical behavior is consistent with the
classical 3D Heisenberg universality class, as expected. We discuss the 
general nature of the transition from quantum mechanical to classical 
(thermal) order parameter fluctuations at a continuous $T_{\rm c} > 0$ 
phase transition.
\end{abstract}

\pacs{PACS numbers: 75.10.Jm, 75.40.Cx, 75.40.Mg, 05.30.-d}

\vskip2mm]

The Heisenberg quantum antiferromagnet \cite{bethe,anderson} is one of the 
most important models in solid state physics. A wide variety of physical 
behavior can be realized, depending on the geometry of the lattice over 
which the pair interaction $J{\bf S}_i \cdot {\bf S}_j$ is summed. On a 1D 
linear chain, the quantum fluctuations destroy the long range order, leading 
to a ground state with gapped (for integer $S$) or critical (for half 
odd-integer $S$) fluctuations \cite{luther,haldane,nums1}. On a 2D square 
lattice the ground state is ordered for all $S$\cite{affleck,reger}. For 
$T > 0$ the order is destroyed by thermal fluctuations \cite{mermin}. In 3D, 
a second-order phase transition to an ordered state at $T_{\rm c} > 0$ is 
expected. The quantum fluctuations of the order parameter become negligible 
(relative to the classical, thermal fluctuations) in the neighborhood of 
$T_{\rm c}$ and the critical behavior should therefore be that of the 
classical 3D Heisenberg model. The quantum fluctuations vanish also in
the limit $T \to \infty$, and hence their relative strength should have
a maximum for $T_{\rm c} < T < \infty$. A universal ``quantum critical'' 
regime \cite{chn,chubukov,sandvik1} is in principle possible if the system 
is sufficiently close to a quantum critical ($T_c=0$) point \cite{sachdev}.

Analytic calculations, employing several different approximations at
finite $T$, have recently been carried out for the $S=1/2$ Heisenberg model
on a 3D simple cubic lattice \cite{approx}.  Critical temperatures in the range
$T_{\rm c}/J = 0.89 - 1.13$ were obtained. Results for thermodynamic 
quantities depend strongly on the approximation used. High temperature 
series expansions have been used for the thermodynamics at high temperatures
\cite{rushbrooke}, but cannot be expected to be accurate close to $T_{\rm c}$.
Quantum Monte Carlo (QMC) methods have been important for non-approximate
calculations for 2D quantum antiferromagnets, but so far only limited 
results have been obtained for 3D systems \cite{qmc3d}. We here present 
detailed QMC calculations aimed at reliably extracting $T_{\rm c}$ as well 
as providing unbiased results for the temperature dependence of thermodynamic
quantities. We also discuss the precise nature of the crossover from quantum
mechanical to purely classical thermal order parameter fluctuations at the
critical point. In particular, we show that the ratio 
$S({\bf Q})/[T\chi ({\bf Q})]$ of the static order parameter structure 
factor and susceptibility obeys a finite-size scaling at $T_{\rm c}$ that is 
{\it completely universal}, independent of the exponents of the phase 
transition or the dimensionality.

For the QMC simulations we have used the Stochastic Series Expansion (SSE) 
algorithm \cite{sse} (a generalization of Handscomb's method \cite{handscomb})
for $L^3$ lattices with even $L$ up to 16 in the grand canonical ensemble. 
The method does not introduce any systematic errors in computed quantities. 
As a significant development 
allowing us to efficiently obtain results on a dense temperature grid, we have
introduced ``tempering'' within the SSE algorithm. In this approach 
\cite{tempering}, the configuration space of the simulation is extended to 
include a range of temperatures between which the system can make transitions,
hence allowing calculations for a large number of temperatures much faster 
than with separate fixed-$T$ simulations. Also crucial in this work is the 
use of covariance effects to increase the numerical precision. The scheme 
introduced in \onlinecite{sandvik2} allows for reduction of the statistical 
errors by almost three orders of magnitude for the 3D Heisenberg model. 

\begin{figure}
\centering
\epsfxsize=6.5cm
\leavevmode
\epsffile{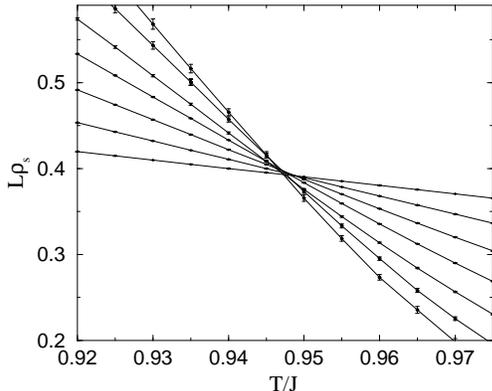}
\caption{Spin stiffness vs.~temperature for systems with $L=4-16$ 
(the slope increases with $L$).}
\label{figrhos}
\end{figure}

In the path integral formulation, a quantum system at finite $T$ corresponds 
to a classical system with an extra ``imaginary time'' dimension of length 
$L_\tau \sim 1/T$ \cite{sondhi}. Standard finite-size scaling arguments
\cite{cardy} can be applied to this system. In order to determine the critical
temperature $T_{\rm c}$, we here first consider the $T$ and $L$ dependence of
the spin stiffness $\rho_{\rm s}$. With the SSE method $\rho_s$ can be 
calculated directly from the global winding number fluctuations in periodic 
systems \cite{note1}, as  described in \onlinecite{sandvik3}. 
Hyperscaling theory predicts the finite-size scaling of the stiffness at 
the critical point \cite{wallin},
\begin{equation}
\rho_{\rm s}(T_{\rm c}) = L^{2-d-z},
\label{rhoscale}
\end{equation}
where $d$ is the dimensionality and $z$ is the dynamic critical
exponent which is 
zero for a $T_c > 0$ transition. For the 3D Heisenberg model, 
$L \rho_s$ graphed versus $T$ for different $L$ should hence intersect at 
$T_{\rm c}$. For $L=4-16$ the curves indeed intersect almost at a single 
point, as shown in Figure \ref{figrhos}. For the smallest systems there is a 
clear tendency of the intersection points for sizes $L$ and $L+2$ to move to
lower $T$ with increasing $L$, but for $L \ge 10$ no shifts can be seen within
the statistical errors. From these results we therefore estimate
$T_{\rm c} \approx 0.945J$.

\begin{figure}
\centering
\epsfxsize=6.8cm
\leavevmode
\epsffile{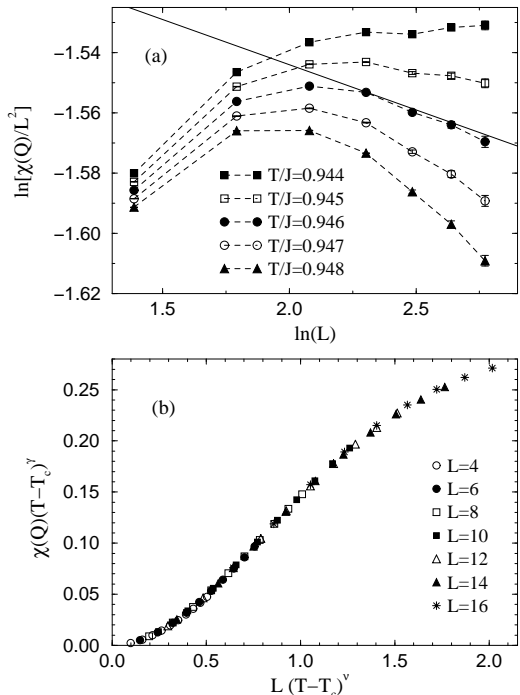}
\caption{(a) Size dependence of the staggered susceptibility 
divided by $L^{2}$ at temperatures close to $T_{\rm c}$. The straight line
is the $L \to \infty$ behavior (slope $-0.03$) expected with the exponent 
$\eta = 0.03$. (b) Finite-size scaling of the staggered 
susceptibility above $T_{\rm c}$, using the classical 3D Heisenberg
exponents $\eta = 0.03$ and $\nu =0.71$, and assuming $T_{\rm c}/J=0.9459$.}
\label{figscale}
\end{figure}

In the thermodynamic limit, the sublattice magnetization $|\langle S^z_i
\rangle |$ (the order parameter) increases continuously from zero for 
$T < T_{\rm c}$. Consider the static structure factor and susceptibility
\begin{mathletters}
\begin{eqnarray}
S({\bf q}) & = & {1\over L^3} \sum\limits_{i,j} 
{\rm e}^{i{\bf q} \cdot ({\bf r}_j - {\bf r}_i)} 
\langle S^z_j S^z_i \rangle,
\label{structq} \\
\chi({\bf q}) & = & {1\over L^3} \sum\limits_{i,j} 
{\rm e}^{i{\bf q} \cdot ({\bf r}_j - {\bf r}_i)} 
\int_0^\beta d\tau \langle S^z_j (\tau) S^z_i (0) \rangle . 
\label{suscq}
\end{eqnarray}
\label{sqxq}
\end{mathletters}
At the ordering wave-vector ${\bf Q}=(\pi,\pi,\pi)$, $S$ and
$\chi$ should both scale with the system size as $L^{2-\eta}$ at 
$T=T_{\rm c}$, where $\eta \approx 0.03$ for the 3D Heisenberg transition 
\cite{peczak}. $\chi ({\bf Q})$ corresponds to the full space-time integral 
of the sublattice magnetization of the $3+1$ dimensional classical system 
and can therefore be expected to be less affected by corrections to scaling.
Figure \ref{figscale}a shows $\ln{(\chi /L^2)}$ vs.~$\ln{(L)}$ at 
temperatures close to the $T_{\rm c}$ estimated above using the stiffness. 
Asymptotically we expect the points to fall on a straight line with slope
$-\eta \approx -0.03$ if $T=T_{\rm c}$, and eventually diverge upward or 
downward below and above $T_{\rm c}$, respectively. At $T \approx 0.946J$ 
the $L \ge 10$ data show the expected scaling. Investigating the behavior
for several temperatures in this neighborhood, and using the estimated
accuracy of the exponent $\eta$ \cite{peczak}, gives $T_{\rm c}/J = 0.946 
\pm 0.001$, in excellent agreement with the spin stiffness scaling.

With $T_{\rm c}$ determined, the expected scaling for $T > T_{\rm c}$ can
be tested. In the thermodynamic limit $\chi$ should diverge as 
$t^{-\gamma}$, where $t=|T-T_{\rm c}|$ and $\gamma = \nu(2-\eta) 
\approx 1.40$ \cite{peczak}.
Finite-size scaling predicts $\chi_L (t) = \chi_\infty (t) 
f[\xi (t)/L]$, where the correlation length $\xi \sim t^{-\nu}$. Hence, 
$\chi_L (t)t^\gamma$ graphed versus $Lt^\nu$ for different $L$ should 
collapse onto a single curve. As shown in Figure \ref{figscale}b, the data 
are indeed collapsed, confirming the expected universality class and the 
estimated $T_{\rm c}$.

\begin{figure}
\centering
\epsfxsize=6.9cm
\leavevmode
\epsffile{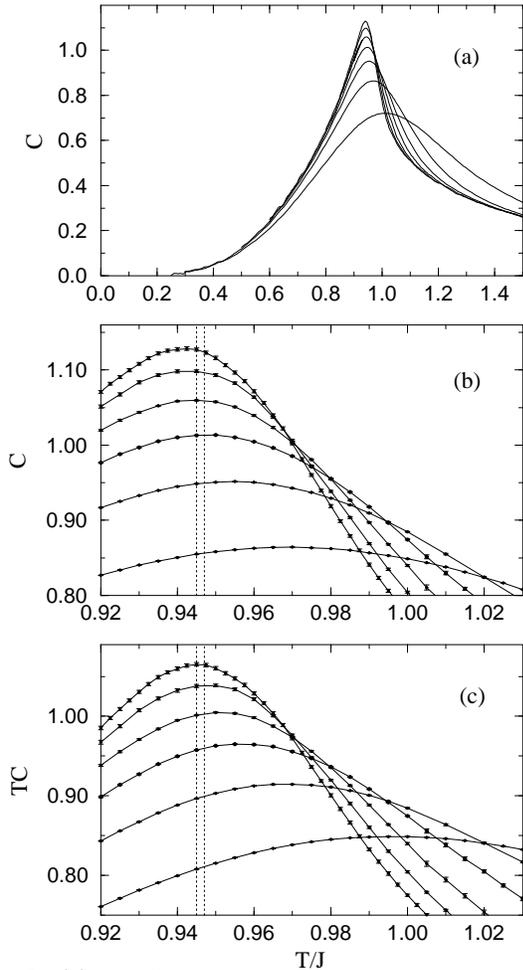}
\caption{(a) Specific heat vs.~temperature for $L=4-16$. The peak height 
increases with $L$. (b) the behavior of the peak for $L=6-16$ on a more 
detailed scale. (c) The peak behavior of $TC$. The estimated $T_{\rm c}$
is between the dashed lines.}
\label{figc}
\end{figure}

We next consider the specific heat, $C={(\partial E / \partial T)/L^3}$.
Figure \ref{figc}a shows the temperature dependence for different system 
sizes. The position of the maximum of $C$ represents a size dependent 
definition of the critical point, and, as shown in more detail in 
Fig.~\ref{figc}b, seems to indicate that $T_{\rm c}$ is lower than the
value estimated above. We believe that the reason for this apparent 
inconsistency is to be found in the fact that changing $T$ of the quantum 
system corresponds to changing $L_\tau$ of the corresponding 3+1 dimensional 
classical system. The calculated $C$ hence corresponds to 
$\partial E / \partial L_\tau^{-1}$. The critical behavior, governed by the 
exponent $\alpha$, of the actual classical specific heat and $C$ of the 
quantum system should nevertheless be the same (except exactly at 
a $T_{\rm c}=0$ quantum critical point) since the boundary between the 
ordered and disordered phases in the classical $(T,L_\tau)$ plane is analytic
for $T > 0$. However, for small systems the shift of the peak position with 
$L$ is relatively large, and the associated $L_\tau$ variation can then cause 
significant deviations from the asymptotic finite-size scaling behavior of 
$C$. In order to partially compensate for the volume change due to the 
variable $L_\tau$, we consider the corresponding quantity defined per full 
space-time volume, i.e., $TC$. Figure \ref{figc}c shows the behavior of 
the peak of $TC$. Compared to the behavior of $C$, we note that the peaks 
for these small systems are shifted noticeably to the right, and now appear
to be consistent with $T_{\rm c} \approx 0.946J$.

For the 3D classical Heisenberg transition $\alpha$ is negative 
($\alpha \approx -0.11$) \cite{peczak}, and hence $C$ should exhibit a 
cusp-like singularity maximum as $L \to \infty$. Finite-size scaling predicts
the size dependence of the maximum value as $C_{\rm max} (L) = C_{\rm max} 
(\infty) - aL^{\alpha/\nu}$ ($\nu \approx 0.71$). The data for $TC$ is indeed
well described by this expression for $L > 6$, and gives 
$T_{\rm c}C_{\rm max}(\infty) \approx 2.4$. The finite-size shift of the peak 
temperature should scale as $L^{-1/\nu}$. The results do not show 
this behavior. It is quite 
likely that this is due to the $L_\tau$ effect discussed above, which is not
fully eliminated in $TC$. We note, however,
that the expected scaling of $C$ is not 
easily observed even for the classical 3D Heisenberg model, for which 
significantly larger systems have been studied \cite{peczak}. A qualitative 
difference is that the peak shifts to higher $T$ with increasing $L$ in the 
classical case \cite{peczak}. 
As discussed above, for the quantum model
the peak in $C$ initially moves with increasing $L$ from above $T_{\rm c}$ 
to slightly below $T_{\rm c}$. As $L \to \infty$, $C$ and $TC$ should both
peak exactly at $T_{\rm c}$. One can therefore expect a change in the sign of 
the $C$ peak shift for some large $L$, and then the same asymptotic behavior 
as for the classical model.

We now discuss the nature of the crossover to purely classical (thermal)
fluctuations of the order parameter as $T \to T_{\rm c}$ from above. For
this purpose we note the well known sumrules
\begin{eqnarray}
S({\bf q}) & = & {1\over \pi}\int_0^\infty d\omega 
(1+{\rm e}^{-\beta\omega}) S({\bf q},\omega), \\
\chi({\bf q}) & = & {2\over \pi}\int_0^\infty d\omega 
\omega^{-1} (1-{\rm e}^{-\beta\omega}) S({\bf q},\omega),
\end{eqnarray}
where $S({\bf q},\omega)$ is the dynamic structure factor, and define the
ratio
\begin{equation}
R({\bf q}) = S({\bf q})/[T\chi({\bf q})].
\end{equation}
For purely thermal
fluctuations of momentum ${\bf q}$, all the spectral
weight is in a $\delta$ function at $\omega=0$, and the classical relation
$R({\bf q})=1$ is obeyed. Quantum $(|\omega| > 0)$ fluctuations increase $R$
above $1$. For the 3D antiferromagnet the fluctuations at the 
antiferromagnetic momentum ${\bf Q}$ should be purely classical for 
$T \le T_{\rm c}$. As $T \to \infty$, $R({\bf q}) \to 1$ for any ${\bf q}$. 
For $T \to T_{\rm c}$ from above, $L_\tau \ll \xi$ and in evaluating
Eqs.~(\ref{sqxq}) the imaginary-time dependence of the long-distance 
correlation function can then be expanded to 
leading order in $\tau/ r$ (or $\tau^{1/z'}/ r$ if the system is 
not Lorenz invariant, which does not change the result). This gives 
$R({\bf Q})-1 \sim \xi^{-2} \sim t^{2\nu}$, and the size dependence at 
$T=T_{\rm c}$ should therefore be (substituting $\xi \to L$):
\begin{equation}
R({\bf Q})-1 \sim L^{-2}.
\label{rscale}
\end{equation}
Note that this scaling does not contain any model-dependent exponents,
and remains valid in any number of 
dimensions (with the $S({\bf Q})$ and $\chi({\bf Q})$ corresponding to 
the relevant order parameter) as long as $T_{\rm c}>0$ and the order parameter
does not commute with the Hamiltonian. (In the commuting case $R({\bf Q})=1$ 
for all $L$ and $T$). $R({\bf Q})$ should therefore in principle be useful, 
as an alternative to the stiffness scaling (\ref{rhoscale}), for extracting 
$T_{\rm c}$ in cases where the critical exponents are not known. 

\begin{figure}
\centering
\epsfxsize=7.2cm
\leavevmode
\epsffile{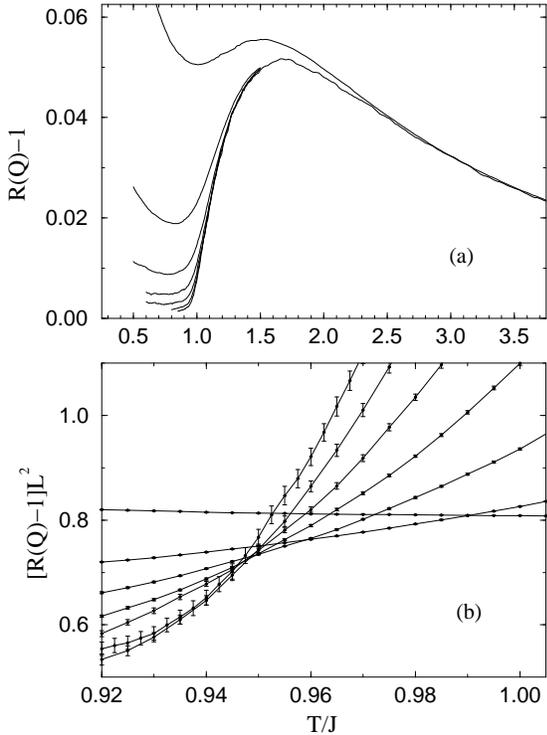}
\caption{(a) Temperature dependence of $R({\bf Q})-1$ for system sizes 
$L=4-16$ (values decrease with increasing $L$). (b) The same quantity 
multiplied by $L^2$ close to $T_{\rm c}$ (slope increases with $L$).}
\label{figratio}
\end{figure}

Figure \ref{figratio}a shows $R({\bf Q})-1$ versus temperature for 
3D Heisenberg lattices. The finite-size gap leads to a $T \to 0$ divergence 
starting at a temperature which decreases with increasing $L$. For the larger
systems there is a maximum at $T \approx 1.7J$, and a rapid drop below 
this temperature. The maximum value $R_{\rm max}-1 \approx 0.05$, which is 
about half of the value for the 2D Heisenberg model \cite{chubukov,sandvik4},
reflecting the reduction of quantum fluctuations with increasing 
dimensionality. 

Figure.~\ref{figratio}b shows the $T$ 
dependence of $[R({\bf Q})-1]L^2$ for different system sizes. The curves 
indeed intersect. The intersection between $L$ and $L+2$ moves to lower $T$ 
with increasing $L$, and is consistent with the estimated $T_{\rm c} 
\approx 0.946J$. The proposed scaling law, Eq.~(\ref{rscale}), for 
$R({\bf Q})$ is hence confirmed.

It is well known that a state of dominant classical order parameter 
fluctuations can extend to a finite regime well above $T_{\rm c}$ 
\cite{chn,sachdev}. This is not in conflict with the universal scaling 
behavior of $R({\bf Q})$ noted here, according to which the classical value
$R({\bf Q})=1$ always is approached algebraically as $T \to T_{\rm c}$.
To see this, consider a system of 2D layers coupled by a very weak 
interlayer coupling $J_\perp$. As the temperature is lowered, the 
correlation length within the layers initially  grows exponentially 
\cite{chn} but there are no significant correlations between the layers. 
For $T$ sufficiently low, but $T \gg T_{\rm c}$, they are in the 
``renormalized classical'' \cite{chn} regime and $R({\bf Q}) \to 1$ 
exponentially \cite{sandvik4}. However, as $T$ is further lowered to the 
point, close to $T_{\rm c}$, where the layers start to correlate with each 
other, there will be a crossover to a $\xi \sim t^{-\nu}$ scaling behavior
and the form (\ref{rscale}) for $R({\bf Q})$. At this point $R({\bf Q})$ 
is already exponentially close to $1$, and the further algebraic
reduction of $R({\bf Q})$ as $T \to T_{\rm c}$ is exponentially small. 

In summary, we have determined the critical temperature $T_{\rm c}/J 
= 0.946 \pm 0.001$ of the 3D Heisenberg antiferromagnet with $S=1/2$, 
using QMC data of high precision. We have discussed the general nature of the 
crossover to purely classical thermal order parameter fluctuations as 
$T \to T_{\rm c}$ from above, and shown that asymptotically the transition 
is universal in the ratio $S({\bf Q})/[T\chi ({\bf Q})]$, independent of 
the universality class of the critical point. We have also noted how quantum 
mechanical effects can affect the finite-size scaling behavior of the 
specific heat.

I would like to thank D. K. Campbell for stimulating discussions and
valuable comments on the manuscript. This work was supported by the NSF under 
grant No.~DMR-97-12765. The numerical simulations were carried out at 
the NCSA.

\end{document}